\documentclass{statsoc}

\usepackage[a4paper]{geometry}
\usepackage[textwidth=8em,textsize=normal]{todonotes}
\usepackage{graphicx}
\usepackage{amsmath}
\usepackage{natbib, float}
\usepackage{amsfonts}
\usepackage{lscape}
\usepackage{color}
\usepackage{multirow}
\usepackage{multicol}
\usepackage{mathtools}
\usepackage{dsfont}

\usepackage{etoolbox}
\usepackage[caption = false]{subfig} 
\usepackage{enumerate}
\usepackage{comment}

\makeatletter
\patchcmd{\@makecaption}
  {\parbox}
  {\advance\@tempdima-\fontdimen2} 
  {}{}
\makeatother

\DeclareMathOperator{\pb}{\mathbf{p}}

\DeclareMathOperator{\Y}{\boldsymbol{Y}}
\DeclareMathOperator{\y}{\mathbf{y}}

\DeclareMathOperator{\bmu}{\boldsymbol{\mu}}

\DeclareMathOperator{\bbeta}{\boldsymbol{\eta}}

\DeclareMathOperator{\btheta}{\boldsymbol{\theta}}
\DeclareMathOperator{\x}{\mathbf{x}}

\DeclareMathOperator{\Ex}{\mathbb{E}}
\DeclareMathOperator{\Vx}{\mathbb{V}}

\DeclareMathOperator{\diag}{\mathrm{diag}}

\DeclareMathOperator{\blambda}{\boldsymbol{\lambda}}

\DeclareMathOperator{\PA}{\mathbb{P}_{\boldsymbol{A} |\blambda}}

\newcommand{\supp}{supplementary materials}

\title[Prior predictive elicitation]{Flexible Prior Elicitation via the \\ Prior Predictive Distribution}

\author[Hartmann, Agiashvili, B\"{u}rkner \& Klami]{Marcelo Hartmann$^{1}$, Georgi Agiashvili$^{1}$, Paul B\"{u}rkner$^{2}$ \& Arto Klami$^{1}$}

\address{Department of Computer Science, University of Helsinki, Helsinki, Finland \ $^{1}$ \\
Department of Computer Science, Aalto University, Espoo, Finland \ $^{2}$}

\email{\\ marcelo.hartmann@helsinki.fi \\ georgi.agiashvili@helsinki.fi \\ paulburkner@gmail.com \\ arto.klami@helsinki.fi}

\begin{document}

\begin{abstract}
  The prior distribution for the unknown model parameters plays a crucial role in the process of statistical inference based on Bayesian methods. However, specifying suitable priors is often difficult even when detailed prior knowledge is available in principle. The challenge is to express quantitative information in the form of a probability distribution. Prior elicitation addresses this question by extracting subjective information from an expert and transforming it into a valid prior. Most existing methods, however, require information to be provided on the unobservable parameters, whose effect on the data generating process is often complicated and hard to understand. We propose an alternative approach that only requires knowledge about the observable outcomes -- knowledge which is often much easier for experts to provide.  Building upon a principled statistical framework, our approach utilizes the prior predictive distribution implied by the model to automatically transform experts judgements about plausible outcome values to suitable priors on the parameters. We also provide computational strategies to perform inference and guidelines to facilitate practical use.
\end{abstract}

\section{INTRODUCTION}

The Bayesian approach for statistical inference is widely used both in statistical modeling and in general-purpose machine learning. It builds on the simple and intuitive rule that allows updating one's {\it prior beliefs} about the state of the world through newly made observations (i.e., data) to obtain {\it posterior beliefs} in a fully probabilistic manner. Nowadays, the Bayesian approach can routinely be used in a vast number of applications due to combination of powerful inference algorithms and probabilistic programming languages \citep{pp:2018}, such as Stan \citep{stan:2019}. 

Despite available computational tools, the task of designing and building the model can still be difficult. Often, the user building the model can safely be assumed to have good knowledge of the phenomenon they are modeling. However, they additionally need to have sufficient statistical knowledge in order to formulate the domain assumptions in terms of probabilistic models which are sensible enough to obtain valid inference. This is by no means an easy task for the majority of users. Hence, the model building process is often highly iterative, requiring frequent modifications of modeling assumptions, for example, based on predictive checks and model comparisons; see \cite{daee:2017}, \citet{schad:2019} and \citet{sarmaetla}  for attempts of formalising the modeling workflow.

We focus on one particular stage of the modeling process, namely the problem of specifying priors for the model parameters. The prior distribution lies at the heart of the Bayesian paradigm and must be designed coherently to make Bayesian inference operational \citep[e.g., see][]{kadanew:1998}. The practical difficulty, though, even for more experienced users, is the encoding of one's actual prior beliefs in form of parametric distributions. The parameters may not even have direct interpretation, and the effect of the prior on the data generating mechanism can be quite involved and show large disparity with respect to what the user's prior beliefs over the data distribution could be \citep{kadane:1980}.

The existing literature addresses this issue via {\it expert knowledge elicitation}. This is understood as the process of extracting the expert's information (knowledge or opinion) related to quantities or events that are uncertain, and expressing them in the form of a probability distribution, the prior. See, for example, the works by \citet{lindley:1983}, \citet{genest:1985}, and \citet{gelfandetal:1995} for early ideas and introduction. See \citet{garthwaite:2005} and \citet{ohagan:2019} for detailed reviews of expert elicitation procedures and guidelines.

The majority of the knowledge elicitation literature is on eliciting information with respect to the {\it parameters} of the model, that is, asking the expert to make statements about plausible values of the parameters. The early works do this within specific parametric prior families, whereas more recently, \citet{ohagan:2004}, \citet{gosl:2005} and \citet{jerem:2007} have proposed nonparametric approaches based on Gaussian processes \citep{ohagan:78}, allowing more more flexibility. Even though the prior itself can be of flexible form, the elicitation process is typically carried out on a parameter-by-parameter basis so that
each parameter receives its own independent univariate prior. As a result, the implied joint prior on the whole set of parameters is often unreasonable.
Although \citet{moala:2010} generalized the approach of \cite{gosl:2005} to multivariate priors, the resulting process is difficult for experts, since they are required to express high-dimensional joint probabilities. Hence, its practical use is basically limited to just two dimensions.

Independently of whether we assign individual or joint priors on the model parameters, any prior can only be understood in the context of the model it is part of \citep[e.g.,][]{Gelman2017, pcpriors}. This point may be obvious but its practical implications are far reaching. Subject matter experts, who may understandably lack in-depth knowledge of statistical modeling, are left with the task of assigning sensible priors on parameters whose scale and real-world implications are hard to grasp even for statistical experts.

For this reason, \cite{kadane:1980} and \cite{akbarov:2009} argue that prior elicitation should be conducted using observable quantities, by asking statements related to the {\it prior predictive distribution}, that is, the distribution of the data as predicted by the model conditioned on the parameters' prior, instead of directly referring to the prior on the unobservable parameters. After eliciting the prior predictive distribution, the information can then be transformed into priors on the parameters by a suitable methodology. The logic of using the prior predictive distribution is that the expert should always have an understanding about plausible values of the observable variables based on their own domain knowledge -- even if they may not fully understand the statistical model and the role of parameters used to represent the underlying data generating mechanism. After all, what is an expert if they do not understand their own data?

From a predictive viewpoint, \cite{kadane:1980}, \cite{kadanew:1998}, \cite{geisse:1993}, and \cite{akbarov:2009} present practical methods for recovering the prior distribution via expert's information on the prior predictive distribution. Those methods are based on specifying particular moments of the prior predictive distribution for a Gaussian linear regression model, or on providing prior predictive probabilities for fixed subregions of the sample space where the prior distribution is assumed to be univariate.  In the latter case,  the strategy is to perform least-squares minimization between theoretical probabilities and those probabilities quantified by the expert. However, in the sense of \cite{ohagan:2004}, these approaches neglect the fact that the expert's information itself can be uncertain and provide no measure for whether the chosen predictive model is able to reproduce the expert's probabilistic judgements well enough. That is to say, existing methods do not take into account imprecisions in probabilistic judgements when constructing the prior predictive distribution, nor do they provide a principled framework which would guide the experts to select a predictive model and/or prior distribution matching their knowledge \citep{jeffreys:1980,winkler1967}.  

Our contribution addresses the question of prior elicitation via prior predictive distributions using a principled statistical framework which 1) makes prior elicitation independent on the specific structure of the probabilistic model from the users' viewpoint, 2) handles complex models with many parameters and potentially multivariate priors, 3) fully accounts for uncertainty in experts/users probabilistic judgements on the data, and 4) provides a formal quality measure indicating if the chosen predictive model is able to reproduce experts' probabilistic judgements. Our work provides both the theoretical basis as well as flexible tools that allow the modeller to express their knowledge in terms of the probability of the data while taking into account the uncertainty in their judgements.

In Section 2, we establish the basic notation and explain why the prior predictive distribution is better suited to represent expert's opinions.  Sections 3 and 4 introduce the methodology to tackle imprecise probabilistic judgements via a principled statistical framework, and general computational procedures to recover the hyperparameters of a prior distribution. The development is interleaved with practical examples illustrating the core concepts and demonstrating its practical use -- via concrete instantiations for multivariate prior elicitation for generalized linear models and a small-scale user study comparing the proposed methodology for classical prior elicitation directly on model parameters. We close the paper in Section 5, where conclusions and potential future directions are presented.

\section{NOTATION AND PRELIMINARIES} 
\label{PrPD}

\subsection{Bayesian approach to Statistical inference}

The process of performing Bayesian statistical inference usually starts by building a joint probability distribution  of observable variables/measurements $\Y$ and unobservable parameters $\btheta$. The corresponding marginal distribution with respect to $\btheta$ is referred to as the prior distribution and the marginal distribution with respect to $\Y$ is referred to as the prior predictive distribution. According to the Bayesian paradigm, the prior distribution should be designed independently of the measurement outcomes, that is to say, it must reflect our prior knowledge about the parameters $\btheta$ before seeing the actual independent measurements $\y_1, \y_2, \ldots$ (i.e., realizations of $\Y$) obtained in the experiments \citep{berger:93, ohagan:94}.
After having obtained the measurements, the posterior distribution of $\btheta$ arises from the joint distribution by conditioning on $\y_1$, $\y_2$, $\ldots$ \citep{ohagan:94}.

\subsection{Prior predictive distribution}

Let $\Y = [Y_1 \ldots Y_S]$ be a $S$-dimensional vector of observable variables and denote the sample space $\Omega$ as a subset of $\mathbb{R}^S$. Hereafter we denote by $\Y|\btheta \sim \pi_{\Y|\btheta}$ our data probability distribution conditioned on the parameters. We also write $\btheta \sim \pi_{\btheta}$ where $\btheta \in \Theta \subseteq \mathbb{R}^D$ and $\pi_{\btheta}$ belongs to a given family of parametric distributions, say $\mathcal{F}_{\blambda}$ indexed by a hyperparameter vector $\blambda$. Then, by marginalizing out the parameters $\btheta$, the prior predictive distribution is given by
\begin{align} \label{eq:mpp}
\pi_{\Y}(\y | \blambda) = \displaystyle\int_\Theta \pi_{\Y |\btheta}(\y|\btheta) \pi_{\btheta}(\btheta| \blambda) \ \mathrm{d} \hspace{-0.03cm} \btheta.
\end{align}

The prior predictive distribution is not to be confused with the marginal likelihood of observed data, which is obtained by marginalization over $\btheta$ of the observed data's sampling distribution times the prior \citep[e.g.,][]{jeffreys:1980}.

Given any subset $A \subseteq \Omega$, the prior predictive probability of $A$, denoted as $\mathbb{P}(\Y \in A| \blambda)$, can be obtained by exchanging the order of integration via the Fubini-Tonelli theorem \citep{folland:2013} as
\begin{align} \label{eq:mpp1}
\mathbb{P}_{A| \blambda} &:= \displaystyle\int_A \pi_{\Y}(\y| \blambda) \ \mathrm{d} \hspace{-0.05cm} \y \nonumber \\
&= \Ex_{\btheta}\big(\mathbb{P}_{\Y|\btheta}({\Y \in A|\btheta})\big).
\end{align}
See \supp \ for details. The hyperparameter vector $\blambda$, which defines a particular prior from the set of all priors $\mathcal{F}_{\blambda}$, 
will be treated as constant. Hence, no prior needs to be assigned to it.
Instead, the values of $\blambda$ will be obtained during the prior predictive elicitation method presented below. 


\section{PRIOR PREDICTIVE ELICITATION}\label{sec:sec_3}

Our approach follows \citet{jerem:2007} and \citet{gosl:2005} by approaching the elicitation process as a problem of statistical inference where the information to be provided by the expert is in the form of probabilistic judgements about the
data. However, the solution itself is novel. From an high-level perspective, our elicitation methodology for any Bayesian model can be summarized as follows: 
\begin{enumerate}
  \item Define the parametric generative model for observable data $\Y$ composed by a probabilistic model conditioned of the parameters $\btheta$ and a (potentially multivariate) prior distribution for the parameters. The prior distribution depends on hyperparameters $\blambda$ essentially defining the prior which we seek to obtain (see Section \ref{PrPD}).
  
  \item Partition the data space into exhaustive and mutually exclusive data categories. For each of these
  categories, ask the expert what they belief is
  the probability of the data falling in that category.
  
  \item Model the elicited probabilities from Step 2 as a function
  of the hyperparameters $\blambda$ from Step 1 while taking
  into account that the expert information is itself 
  of probabilistic nature and has inherent uncertainty.
  
  \item Perform iterative optimization of the model from Step 3 to obtain an estimate for $\blambda$ describing the expert opinion best within the chosen parametric family of prior distributions.
  
  \item Evaluate how well the predictions obtained from the optimal prior distribution of Step 4 can describe the elicited expert opinion.

\end{enumerate}

In the remainder of this section, we first introduce the basic formalism for modelling the users' beliefs in Section~\ref{sec:model}, provide a key consistency result in Section~\ref{sec:consistency}, then demonstrate how it can be applied to predictive problems in Section~\ref{sssec:covs}, and finally discuss the interfaces for the actual knowledge elicitation procedure in Section~\ref{sec:elicitation}. Each part is concluded by an example illustrating the concept.

\subsection{Modelling expert opinions}
\label{sec:model}

Our assumption is that the output elicitation procedure provides information as probabilistic assignments regarding the data vector $\Y$ falling within a fixed set of mutually exclusively and exhaustive events $\mathbf{A}$. Such collection of assignments can be considered as the data available for inferring the prior, and is not to be confused by actual measurement data following the generative model. Our focus here is in the mathematical machinery required for converting this information into prior distributions, not taking any stance on how the information is collected from the expert. However, we will briefly discuss the elicitation process itself in Section~\ref{sec:elicitation}.

Let $\mathbf{A}$ $=$ $\{A_1, \ldots, A_n\}$ be a partition of the sample space $\Omega$. Throughout the elicitation procedure, the expert supplies their expected opinions regarding the quantities $\mathbb{P}_{A_i| \blambda}$ for all $i = 1, \ldots, n$. The expert's judgements themselves are not fully deterministic and retain some uncertainty. Also, the expert may be more comfortable to make statements for certain partitions of $\Omega$ than for others. 

To account for the uncertainty in the probability quantifications of $\mathbb{P}_{A_i| \blambda}$, we assume that the obtained judgements $\pb$ follow a Dirichlet distribution \citep{ferguson1973} with base measure given by the prior predictive probabilities $\mathbb{P}_{A_i| \blambda}$ and precision parameter $\alpha$. Hence, for any chosen partition $\mathbf{A}$ of size $n$, we denote the distribution of $\pb$ as
\begin{align} \label{eq:randP}
\pb | \alpha, \blambda \sim \mathcal{D}(\alpha, [\mathbb{P}_{A_1| \blambda} \cdots \mathbb{P}_{A_n| \blambda}]),
\end{align}
where $\mathcal{D}(\cdot)$ stands for Dirichlet distribution and whose multivariate density function reads
\begin{align} \label{eq:dirich3}
\mathcal{D}(\pb|\alpha, \blambda) &= \dfrac{\Gamma(\alpha)}{\prod_{i = 1}^n \Gamma(\alpha \hspace{0.08cm} \mathbb{P}_{A_i| \blambda})} \prod_{i = 1}^n p_i^{\alpha \mathbb{P}_{A_i| \blambda} - 1}.
\end{align}
Naturally, we require $\sum_{i = 1}^n \mathbb{P}_{A_i| \blambda} = 1$.
The Dirichlet density \eqref{eq:dirich3} accounts for the uncertainty inherent to the numerical quantification of the probability vector $\pb$ due to, for example, biases introduced through the mechanisms of elicitation processes (the way in which questions are made), practical imperfection (imprecision) of experts' judgements in probabilistic terms or poor judgements on the effect of parameters in the output model. For details and in-depth discussion, see \cite{ohagan:2004}, \cite{ohagan:2019} and \cite{sarmaetla} .

The hyperparameter $\alpha$ measures how well the prior predictive probability model is able to represent (or reproduce) the probability data provided in the elicitation process.  The larger the values of $\alpha$, the less variance around the expected value $\mathbb{P}_{A_i| \blambda}$. For practical use of this principle, we can find the maximum likelihood estimate (MLE) $\hat{\alpha}$ of $\alpha$, which can be directly understood in terms of the deviance between the prior predictive probability and the experts opinion. More specifically, we have
\begin{align} 
\label{alpha-KL}
\hat{\alpha} \approx \dfrac{n/2 - 1/2}{{\rm KL}(\PA \ || \ \pb)}
\end{align}
where $\PA = [\mathbb{P}_{A_i| \blambda} \cdots \mathbb{P}_{A_n| \blambda}]^\top$ and ${\rm KL}(\PA \ || \ \pb)$ is the Kullback-Leibler divergence between the two distributions. The practical interpretation is that for small KL values, we would not be able discriminate the prior predictive probability from the probability data provided by the expert. See \supp \ for the proof of Equation \eqref{alpha-KL}.

\begin{figure}[t!]
  \includegraphics[scale=0.52]{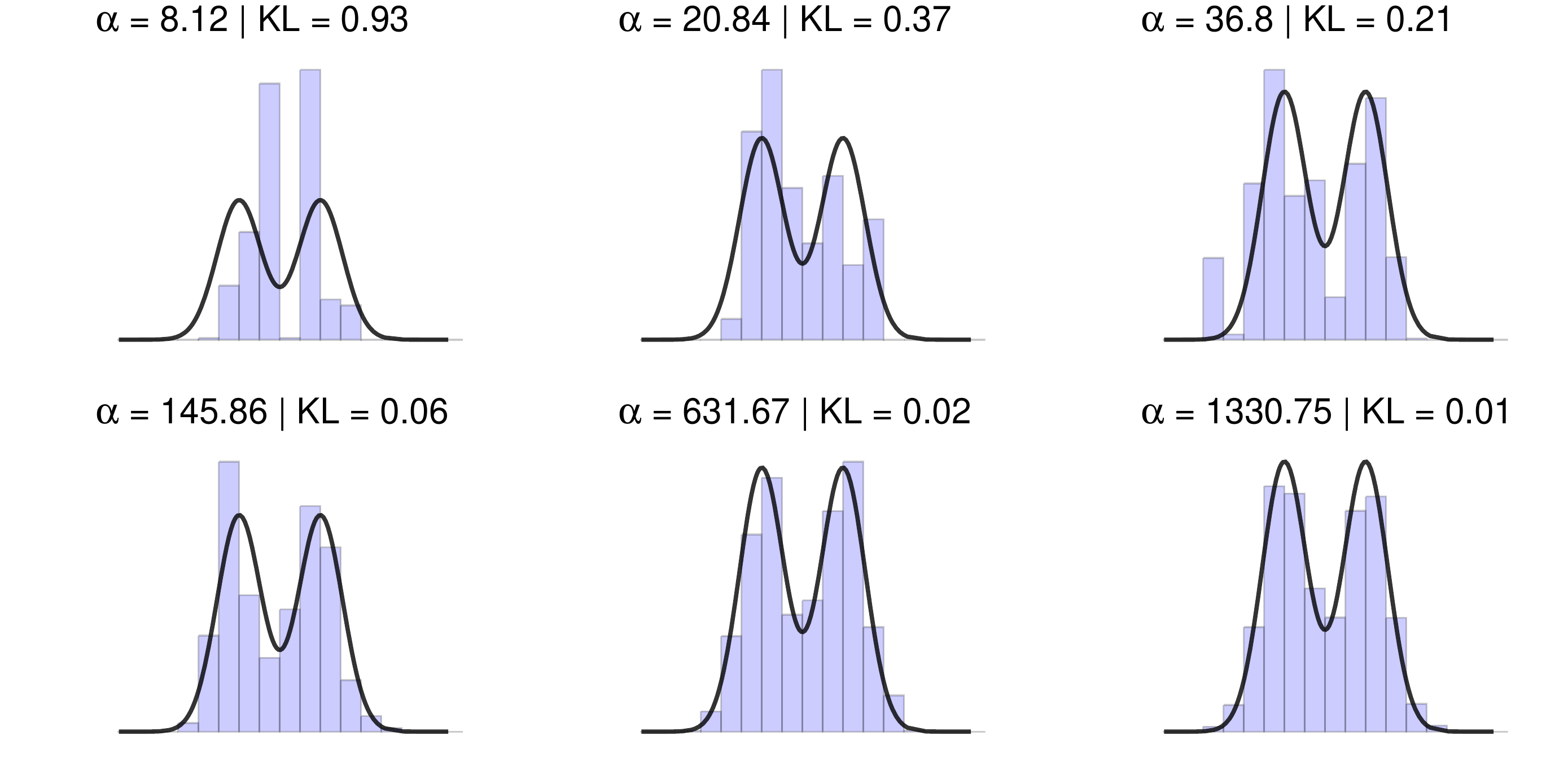}
  \caption{Illustration of the role of the concentration parameter $\alpha$. Large values correspond to scenarios where the prior predictive distribution (solid line) is able to represent expert's opinions (bars) accurately. That is, $\alpha$ provides an accuracy diagnostic for our method with higher values indicating higher accuracy.}
  \label{fig:fig}
\end{figure}

\paragraph{Example:} Consider a generative model given by $Y|\theta \sim \mathcal{N}(\theta, \sigma^2)$ and $\theta \sim \tfrac{1}{2}\mathcal{N}(\mu_1, \sigma_1^2) + \tfrac{1}{2} \mathcal{N}(\mu_2, \sigma_2^2)$. This yields the prior predictive distribution $Y$ $\sim$ $\tfrac{1}{2}\mathcal{N}(\mu_1, \sigma^2 + \sigma_1^2)$ $+$ $\tfrac{1}{2} \mathcal{N}(\mu_2, \sigma^2 + \sigma_2^2)$ with hyperparameters $\blambda$ $=$ $[\mu_1, \ \mu_2, \ \sigma^2, \ \sigma^2_1, \ \sigma^2_2]^\top$. For a set $A =(a, b] \subset \mathbb{R}$, the prior predictive probability is $\mathbb{P}_{A| \blambda}$ $=$ $\sum_{k = 1}^2 \tfrac{1}{2} \Phi\big((a - \mu_k)/\sqrt{\sigma^2 + \sigma^2_k}\big) - \tfrac{1}{2} \Phi\big((b - \mu_k))/\sqrt{\sigma^2 + \sigma^2_k}\big)$. Figure~\ref{fig:fig} illustrates the effect of the $\alpha$ parameter for a given partition $\mathbf{A}$ with $n=10$. For each $\alpha \in \{1, 15, 50, 100, 300, 1000\}$. we generated $\pb$ by sampling from \eqref{eq:randP}, using fixed hyperparameter values of $\mu_1$ $=$ $-\mu_2$ $=$ $2$ and $\sigma^2$ $=$ $\sigma^2_1$ $=$ $\sigma^2_2 = 1$.
  
\subsection{Consistency with respect to partitioning}
\label{sec:consistency}

Even though we work in a Bayesian context looking to recover a prior distribution, the core procedure of our method applies classical statistical inference. Given a numerical vector of probabilities from the elicitation process, the goal is to show that we are able to find the value of certain parameters (in this case the hyperparameters $\blambda$ and concentration $\alpha$ parameter) of the Dirichlet probabilistic model \eqref{eq:randP} which would have most likely generated this particular data (of user's subjective knowledge). In other words, we are aiming to obtain the maximum likelihood estimator (MLE).

To study the MLE, we consider the limit where the partitioning is made increasingly more fine grained by increasing $n$ towards infinity. However, we still only obtain information from the user once (i.e., for a single partitioning). That is, the user is providing more and more information about the probabilities, but does not repeat the procedure multiple times. As we will show below, the MLE is consistent under these circumstances, providing the true $\blambda$ when $n \rightarrow \infty$, under reasonable assumptions.

Recall that equations \eqref{eq:randP} and \eqref{eq:dirich3} represent the probabilistic model of $\pb$ conditioned on the parameters $\bbeta = (\blambda, \alpha)$. Suppose the implied true prior distribution of the expert has hyperparameter values $\blambda_0$ and denote $\bbeta_0 = (\blambda_0, \alpha_0)$. Take the size of the partition $n$ to be large and denote the log-likelihood as $T_{\bbeta}(\pb) = \log \mathcal{D}(\pb|\alpha, \blambda)$ with expectation $Q_{\bbeta_0}(\bbeta) = \Ex_{\mathcal{D}}(T_{\bbeta}(\pb))$. 

We show that the expected log-likelihood is maximized at $\bbeta_0$. By Jensen's inequality, we know that
\begin{align}
\Ex_{\mathcal{D}} & \left[ -\log \dfrac{\mathcal{D}(\pb|\alpha, \blambda)}{\mathcal{D}(\pb|\alpha_0, \blambda_0)} \right] > - \log \Ex \left[\dfrac{\mathcal{D}(\pb|\alpha, \blambda)}{\mathcal{D}(\pb|\alpha_0, \blambda_0)} \right] = 0,
\end{align}
yielding
\begin{align}
 Q_{\bbeta_0}(\bbeta_0) = \Ex_{\mathcal{D}}(T_{\bbeta_0}(\pb)) > \Ex_{\mathcal{D}}(T_{\bbeta}(\pb)) = Q_{\bbeta_0}(\bbeta), \nonumber 
\end{align}
which holds for all $\bbeta$. The expectation $\Ex_{\mathcal{D}}(\cdot)$ is taken with respect to the distribution \eqref{eq:dirich3}. The technical condition to ensure uniqueness of the MLE is that the probabilistic model \eqref{eq:dirich3} must be identifiable\footnote{In practise, this may not be an issue when fitting the model. However, we believe it is important to understand the theoretical properties of the inference process so that we can avoid problems in the optimisation procedures.}. That is, equality of likelihoods must imply equality of parameters: $\mathcal{D}(\pb|\alpha_1, \blambda_1)$ $=$ $\mathcal{D}(\pb|\alpha_2, \blambda_2) \Rightarrow  \bbeta_1 = \bbeta_2$ for all $\pb$. Otherwise we may encounter multiple maxima and thus the prior distribution in the set $\mathcal{F}_{\blambda}$ is not unique. 

\paragraph{Example:} Extending the earlier example, consider a more general generative model where the prior distribution is now $\theta \sim w_1\mathcal{N}(\mu_1, \sigma_1^2) + w_2 \mathcal{N}(\mu_2, \sigma_2^2)$ yielding the prior predictive distribution $Y$ $\sim$ $w_1\mathcal{N}(\mu_1, \sigma^2 + \sigma_1^2) + w_2 \mathcal{N}(\mu_2, \sigma^2 + \sigma_2^2)$, where $w_1$ and $w_2$ are weights summing up to 1 and the hyperparameters  are given by $\blambda$ $=$ $[\mu_1,\mu_2,$ $\sigma^2, \sigma_1^2,$ $\sigma^2_2, w_1, w_2]$. 

\begin{figure}[t!]
  \setlength{\parindent}{-0.45cm}
  \includegraphics[scale=0.115]{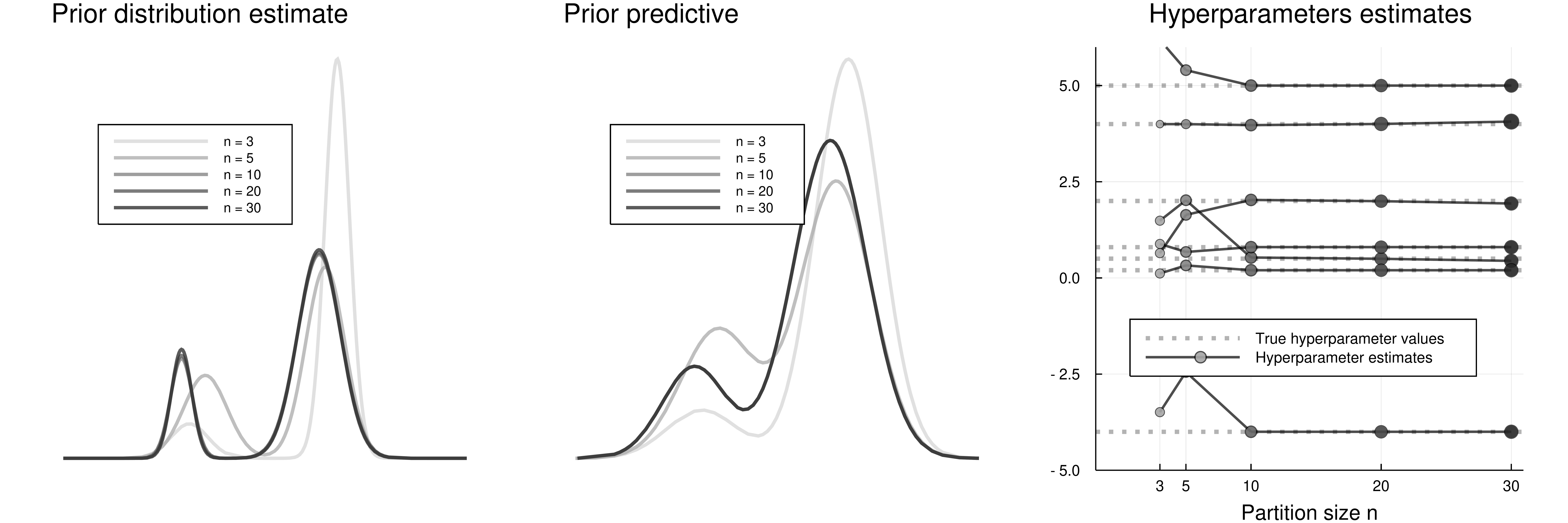}
  \caption{Consistency of the MLE for $\blambda$. {\bf On the right:} 
    All six hyperparameter values converge to the true values as the number of partitions $n$ increases (each line corresponds to one hyperparameter), here converging already roughly for $n = 10$. {\bf On the left and middle:} Both the estimated prior distribution (left) and the corresponding prior predictive distribution (right) converge towards the respective true distributions, depicted as black lines.
    \label{fig:fig_1}}
\end{figure}

Suppose $\alpha$ is fixed and the true prior distribution has hyperparameters $\blambda_0$. We run an experiment where probability vectors are generated from \eqref{eq:randP} with increasing partition sizes. Figure~\ref{fig:fig_1} shows that, as the partition size increases, the estimates $\hat{\blambda}$ converge to $\blambda_0$, which means the prior distribution is recovered from single-sample elicitation of probability data.

\subsection{Covariate-dependent models and multivariate priors} \label{sssec:covs}

Next, we demonstrate how the proposed approach can be used for concrete modelling problems, by detailing the procedure for the widely-used family of generalized linear models \citep[GLM; ][]{glm}. As GLMs typically have several parameters -- one parameter per predicting covariate plus an intercept and potentially a dispersion parameter -- direct specification of the parameters' joint prior is often difficult. However, our prior predictive approach can handle this situation elegantly.

In case of a GLM, our elicitation method requires the selection of sets of covariate values for which the expert is comfortable to express probability judgements about plausible realizations of $\Y$. More formally, for each set of covariates $\x_j$ $=$ $[x_{j, 1} \cdots x_{j, C}]$, $j = 1, \ldots, J$, the expert provides probability judgements $\pb_j = [p_{j, 1} \cdots p_{j, n_j}]$ with $\sum_{i_j = 1}^{n_j} p_{j, i_j} = 1$, where $n_j$ is the partition size for covariate set $j$ implying the partition $\mathbf{A}_j = \{A_{j, 1},$ $\ldots$, $A_{j, n_j} \}$. Under the assumption of the judgement $\pb_j$ being pairwise conditionally independent, we can express the likelihood function of $\alpha$ and $\blambda$ as
\begin{align} \label{eq:dirich4}
\mathcal{D}(\pb_1, \ldots, \pb_J|&\alpha, \blambda) = \dfrac{\Gamma(\alpha)^J} {\prod\limits_{j = 1}^J \prod\limits_{i_j = 1}^{n_j} \Gamma(\alpha \hspace{0.08cm} \mathbb{P}_{A_{j, i_j}| \blambda, \x_j})} \prod_{j = 1}^J \prod_{i_j = 1}^{n_j} p_{j, i_j}^{\alpha \mathbb{P}_{A_{j, i_j}| \blambda, \x_j} - 1} 
\end{align}
where $\mathbb{P}_{A_{j, i_j}| \blambda, \x_j}$ is the prior predictive probability for the set $A_{j, i_j}$ related to covariate set $\x_j$.

Importantly, there is no need for the partitions themselves or their size to be the same throughout the sets of covariate values: For each $j$, the expert can create any partition they are most comfortable with making judgements about. This feature provides much more freedom to the expert in expressing their knowledge of the data compared to alternative methods. For example, to obtain a prior distribution for logistic regression model, the method of \citet{bedrick:1997} requires the user to provide a fixed number of probabilities just enough to make the Jacobians appearing in their method invertible.

\paragraph{Example:}Here we consider a generative model for binary data in the presence of a vector of covariates.
The observable variable conditioned on the parameters is distributed according to a Bernoulli model and we take a multivariate Gaussian distribution as the prior distribution for the vector of parameters in the predictor function. This can be formalized as
\begin{align}
Y|\btheta &\sim \mathcal{B}\big( \Phi(\x^\top \btheta) \big) \ \ \ \ \ \ \ \ \btheta \sim \mathcal{N}_D( \bmu, \Sigma ) 
\end{align}
yielding the prior predictive distribution
\begin{align}
Y \sim \mathcal{B}\Big(p(\x, \blambda) \Big)
\end{align}
with $p(\x, \blambda) = \Phi(\x^\top \hspace{-0.05cm} \bmu/ \sqrt{1 + \x^\top \Sigma \ \x}\hspace{0.05cm})$.

The notation $\mathcal{N}_D(\cdot, \cdot)$ stands for a $D$-dimensional Gaussian distribution and $\mathcal{B}(\cdot)$ for the Bernoulli distribution. The hyperparameter vector $\blambda$ $=$ $[\bmu, \Sigma]$, consists of the prior means $\bmu$ $=$ $[\mu_1, \cdots, \mu_D]$ and prior covariance matrix $\Sigma$. We fix the partitioning throughout the covariate set as $A_{j ,1} = \{ 0 \}$, $A_{j, 2} = \{ 1 \}$ since $Y \in \Omega = \{0, 1\}$. Equation \eqref{eq:mpp1} simplifies to $\mathbb{P}_{A_1| \blambda} = 1 - p(\x, \blambda)$ and $\mathbb{P}_{A_2| \blambda} = p(\x, \blambda)$. 

The parametrisation of the covariance matrix follows the separation strategy suggested by \citet{barnard} on an unconstrained space as presented by \cite{kuro:2003}. That is, the covariance matrix is rewritten as $\Sigma$ $=$ $\diag(\sigma_1^2, \ldots, \sigma_D^2)$ \ $R$ \ $\diag(\sigma_1^2, \ldots, \sigma_D^2)$ where $(\sigma_1^2, \ldots, \sigma_D^2)$ are the variances and $R$ is the correlation matrix.

In the simulation experiment, we vary the dimension $D \in \{2, 3, 4, 5 ,6\}$ and the number of sets of covariates $ J \in \{3, 5, 15, 30, 80\}$. For each $D$ we randomly pick a true value for $\blambda$, and for each covariate set, we draw random probabilities of success/failure from the Dirichlet probability model. Hence, the likelihood is given by \eqref{eq:dirich4}. We repeat the procedure for each $D$ and $J$ where the hyperparameters $\blambda$ are fixed with respect to $J$.

\begin{figure}[t!]
\begin{center}
  \includegraphics[scale=0.45]{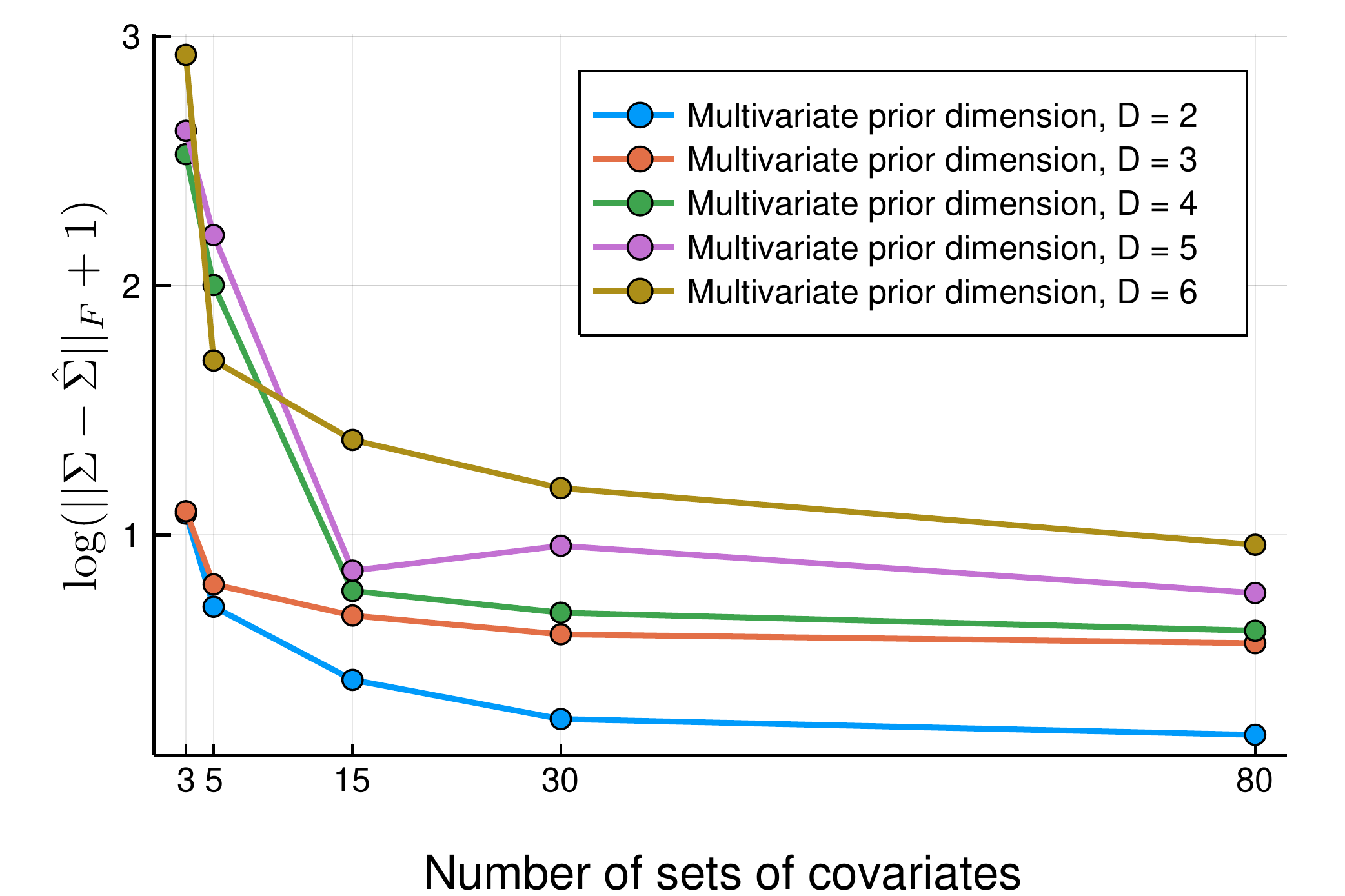}
 \end{center}
  \caption{Convegence of the covariance matrix estimates for multivariate prior elicitation for binary linear regression as a function of the number of covariates $J$ for which the user provides probability estimates, measured using the logarithm of the Frobenius norm of the difference between the true covariance matrix and the estimate. The coloured lines refer to the dimensionality $D$ of the prior distribution, showing that we can effectively elicit multivariate priors of reasonable dimensionality, with naturally increasing difficulty for larger $D$.}
    \label{fig:fig_2}	
\end{figure}

To show the convergence with respect to the estimates of $\Sigma$ obtained from the expert judgements, we compare the logarithm of the Frobenius norm between the estimated covariance matrix and the true covariance matrix (Fig.~\ref{fig:fig_2}). For sufficiently large $J$, roughly from $J=15$ onwards, we are able to accurately elicit multivariate priors up to 5-6 dimensional priors -- this is a significant improvement over earlier methods that have been limited to univariate or at most bivariate priors \citep{moala:2010}. For increasing $D$ from $2$, $3$, $4$, $5$ to $6$, the respective number of hyperparameters in the vector $\blambda$ becomes $5$, $9$, $14$, $20$ to $27$, explaining the increased elicitation difficulty for large $D$.

\subsection{Prior elicitation in practice}
\label{sec:elicitation}

Using the machinery above requires obtaining the probability judgements $\pb$ from the user. The method itself is general, and can be used as part of any practical Bayesian modelling workflow when linked to any particular elicitation interface. We have implemented an extension of the SHELF interface \citep{oakleyshelf:2019} as a reference, by replacing the direct parameter elicitation components with variants that query the user for the prior predictive probabilities. This readily provides practical elicitation methods for the user to specify probabilities by utilizing probability quantiles or roulette chips. This means that probability ratios for events are provided and then individual probabilities are recovered under the natural constraint $\sum_{i = 1}^n p_i = 1$. Hence, the user can choose the way of providing information they feel most comfortable with. Besides graphical interfaces, the elicitation can be carried out by the modeller interviewing a domain expert. Experienced modellers may also choose to simply express some particular priors via providing $\pb$ while designing the model.

\paragraph{Example:} To evaluate the applicability of our
method in practice, we conducted a small user study of $N = 5$ doctoral students of computer science with reasonable statistical knowledge. The task was to elicited priors 
of a human growth model \citep[see][model 1, Section 2]{preece:1978} with a six-dimensional hyperparameter vector $\blambda$. We queried the users for $n_j = 6$ probabilities and $J = 4$ covariates, each corresponding to stature distribution of males at the age of $t \in \{0, 2.5, 10, 17.5 \}$ years. We chose this model because everyone can be expected to have a rough understanding of the observed data and hence can act as an expert. As a baseline, we used a standard elicitation procedure which queries the prior distributions for each parameter directly (again with $n = 6$). Some of these parameters are intuitive (e.g., stature as adult) while some control the quantitative behaviour of the model in a non-trivial way. The model was implemented in \texttt{brms} \citep{burkner:2017} to demonstrate compatibility with existing modelling tools. Gradient-free optimization (see next section) was used for converting the elicited probabilities into priors. Table \ref{tbl:userstudy} shows exemplary for one user how the prior predictive distribution corresponding to $\blambda$ elicited with the proposed method matches well with results of \citet{preece:1978}. When applying direct parameter elicitation, the match was clearly worse because the user was unable to provide reasonable estimates for parameters without an intuitive meaning, despite being provided an explanation of the model and its parameters. In a standardized interview, all users reported that they were more comfortable providing probability judgements for the observables than for the parameters, and that they were more confident that the resulting prior matches their actual subjective prior. See \supp \ for details of the model and user study, as well as results for all users.

\begin{table}
\caption{Result of a real prior elicitation experiment for one user, characterized by statistics of the prior distribution. The proposed approach (Predictive) better matches the parameters  found by fitting the model to actual data \citet[Reference;][]{preece:1978}, compared to direct parameter elicitation (Parametric). This is visible in the lower $\alpha$ estimate as well. The reference column excludes $b$ due to their use of a non-probabilistic model.}

\begin{tabular}{crrrrr}
          &           & \multicolumn{2}{c}{\textbf{Predictive}} & \multicolumn{2}{c}{\textbf{Parametric}} \\ 
Parameter	& Reference & $\Ex[\cdot]$ & $\Vx(\cdot)$ & $\Ex[\cdot]$ & $\Vx(\cdot)$ \\ \hline
$h_1$       & 174.6     & 174.5		   & 0.8                & 176.2  		& 105.3               \\
$h_{t_*}$   & 162.9     & 162.8 	       & 4.2                & 129.1  		& 33.6                \\
$s_0$       & 0.1       & 0.1    	   & $<$ 0.1	 		   & 1.2     		& 1.13                 \\
$s_1$       & 1.2       & 3.3    	   & 0.21               & 1.2     		& 1.13                 \\
$t_*$       & 14.6      & 13.4   	   & 0.01               & 12.5   		& 0.57                 \\
$b$         & $-$          & 15.79  	   & 12.9               & 1.97    		& 4.57                \\
\hline
$\alpha$    & $-$          & 6.9		   & $-$               & 1.2    		& $-$                
\end{tabular}
   \label{tbl:userstudy}
\end{table}

\section{ON THE LEARNING ALGORITHMS}
\label{sec:algoritms}

Having characterized the problem itself and its asymptotic properties, we now turn our attention to the computational problem of estimating the hyperparameter vector $\blambda$ and the uncertainty parameter $\alpha$ in practice. We start by mentioning basic notions for the type of models and properties over which our method is able to accommodate and systematise general purpose model independent computer algorithms.

The methodology presented in Section \ref{sec:sec_3} supports both discrete and continuous components in the observables variables $\Y$, or combinations of both. It also works for any data dimension $S$ and any parameter dimension $D$. Interesting cases are when $S = 1$ and $D > 1$, meaning that, as we have showed previously, we can recover a multivariate prior distribution from probability judgements of $1$-dimensional observable variable. This is novel in the recent literature. 

For arbitrary $S$, where we would possibly work with a multivariate distribution over a vector of observable variables, probabilities for a generic rectangular set $A = \bigtimes_{s = 1}^S (a_s, b_s]$ can be formulated via the cumulative distribution function of the prior predictive distribution \eqref{eq:mpp} as follows. Let $I = (a, b]$ be an interval, $g$ some function with $g : \mathbb{R}^S \rightarrow \mathbb{R}$, and $\Delta^{s}_I$ the difference operator with $\Delta^{s}_I = g(y_1, \ldots, y_{s-1}, b) - g(y_1, \ldots, y_{s-1}, a)$. Then, equation \eqref{eq:mpp1} takes the general form
\begin{align}\label{eq:Fmpp}
\mathbb{P}_{A| \blambda} &= \displaystyle\int_{a_1}^{b_1} \cdots \displaystyle\int_{a_S}^{b_S} \pi_{\Y| \blambda}(y_1, \ldots, y_S) \mathrm{d} y_1  \ldots \mathrm{d} y_S \hspace{-0.04cm}  \nonumber \\ 
&= \Delta^{1}_{I_1} \Delta^{2}_{I_2} \cdots \Delta^{S}_{I_S} F_{_{\Y| \blambda}}(y_1, \ldots, y_S),
\end{align}
where $F_{_{\Y| \blambda}}(\cdot)$ is the cumulative distribution function of the prior predictive distribution \eqref{eq:mpp}. Cases in which $S > 1$ appear, for example, in lifetime analysis or Markovian models. In lifetime analysis, components of electronic equipments are dependent and there is a need to consider bivariate models in the first level of the generative model \citep{lawless2011}. Markovian models are widely used to model natural phenomena such as population growth, climate, traffic, and language models in which multiple measurement variables naturally occur \citep{kijima1997}.

\paragraph{Natural gradients for closed-form cases:}
If equation \eqref{eq:Fmpp} is available in closed-form, usual gradient-based optimisation algorithms are applicable. We recommend using natural gradients \citep{amari:1998}, which have been widely applied for statistical machine learning problems \citep[e.g., see][]{benmark11, salimbeni:2018}. In this case, the Fisher information matrix for $\blambda$ can be computed in closed-form using results from the original parametrisation of the Dirichlet distribution \citep{ferguson1973} as
\begin{align}  \label{eq:fisher}
H_{\blambda} = (\tfrac{d}{d\blambda} \PA)^\top \ H_{\PA} \ (\tfrac{d}{d\blambda} \PA),
\end{align}
where $H_{\PA} = \alpha^2 (\diag(\psi'(\alpha \PA) ) - \psi'(\alpha) \mathds{1} \mathds{1}^\top)$ is the Fisher information matrix of the standard Dirichlet distribution, $\PA$ $=$ $[\mathbb{P}_{A_1| \blambda} \cdots \mathbb{P}_{A_n| \blambda}]^\top$, and $\tfrac{d}{d\blambda} \PA$ $=$ $\big[\tfrac{d}{d\lambda_1} \PA$ $\cdots$ $\tfrac{d}{d\lambda_M} \PA \big]^\top$. The function $\psi'(\cdot)$ is the the derivative of the digamma function and $\tfrac{d}{d\lambda_M} \mathbb{P}$ is the derivative of the vector $\mathbb{P}$ with respect to an element in the vector of hyperparameters $\blambda$. Due to the closed-form expression, we can use natural gradients with almost no additional computational cost. The only extra step is the calculation of $\tfrac{d}{d\lambda_M} \mathbb{P}$ which can be obtained easily with automatic differentiation regardless of the chosen generative model.

\paragraph{Stochastic natural gradients optimization:}
If $\eqref{eq:Fmpp}$ cannot be expressed in closed-form but the equation \eqref{eq:dirich3} or \eqref{eq:dirich4} are differentiable with respect to $\blambda$,
one can use gradient-based optimization with {\it reparametrisation gradients} and automatic differentiation. The elements of $\mathbb{P}$ are expected values with respect to the prior distribution \eqref{eq:mpp1}, and the goal is then to find a pivotal function for the prior \citep[see][page 427, Section 9.2.2]{casella:2001} and obtain Monte-Carlo estimates of it (which is not difficult once we can use the representation \eqref{eq:Fmpp}) and gradients $\tfrac{d}{d\lambda_M} \mathbb{P}$ with very low computational cost according to \cite{figurnov:2018}.

When the generative model has a higher level hierarchical structure, such as $\Y|\btheta_1 \sim \pi(\y|\btheta_1)$, $\btheta_1|\btheta_2 \sim \pi(\btheta_1|\btheta_2)$, $\ldots$, $\btheta_L|\blambda \sim \pi(\btheta_L|\blambda)$, we can show that the elements of $\PA$ and $\tfrac{d}{d\lambda_M} \PA$ can also be computed efficiently together with a stochastic estimation of the hyperparameters' Fisher information matrix. That is
\begin{align}
\PA = \Ex_{X_L} \left( \Ex_{X_{L-1}} \cdots \left(\Ex_{X_1} \left( \mathbb{P}_{\boldsymbol{A}|f_1(\blambda)} \right) \right) \right) 
\end{align}
where $X_\ell$ are pivotal quantities with respect to distributions $\pi(\btheta_\ell|\btheta_{\ell + 1})$ for $\ell = 1, \ldots, L$ and $f_1(\blambda)$ is a function which depends only on the hyperparameters $\blambda$. Gradients are estimated similarly as 
\begin{align}
\dfrac{\mathrm{d}}{\mathrm{d}\lambda_m} \PA &= \Ex_{X_L} \Bigg( \Ex_{X_{L-1}} \cdots  \left(\Ex_{X_1} \left(  \dfrac{\mathrm{d} f_1}{\mathrm{d} \lambda_m} \dfrac{\mathrm{d}} {\mathrm{d} f_1} \mathbb{P}_{\boldsymbol{A}|f_1(\blambda)} \right) \right) \hspace{-0.05cm} \Bigg) 
\end{align}
The equations above can be plugged into \eqref{eq:fisher} to obtain an estimation for the hyperparameters' Fisher information matrix. The proof and detailed explanations are provided in the \supp.

\paragraph{Gradient-free optimization:}
Finally, for completely arbitrary models, we can step outside of gradient-based optimization and use general-purpose global optimization tools for determining $\blambda$. Methods such as Bayesian optimization and Nelder-Mead only require the ability to evaluate the objective \eqref{eq:Fmpp}, and many practical optimization libraries (e.g. \texttt{optimR}) provide extensive range of practical alternatives. For models with relatively small number of hyperparameters, we have found such tools to work well in practice. However, whenever either of the gradient-based methods described above is applicable, we recommend using them due to substantially improve efficiency.

\paragraph{Optimization of $\alpha$:}

Finally, besides $\blambda$, we usually want to estimate $\alpha$ as well which quantifies the uncertainty as explained in Section~\ref{sec:model}. One can either directly optimise \eqref{eq:dirich3} for $(\blambda, \alpha)$ together, or switch optimisation of \eqref{eq:dirich3} for $\blambda$ with fixed $\alpha$ with optimization of \eqref{eq:dirich3} for $\alpha$ with fixed $\blambda$. This may be easier since we have an approximate closed-form expression for $\alpha$ provided in the \supp.

\section{DISCUSSION AND CONCLUSIONS}

Prior elicitation is an important stage in the Bayesian modeling workflow \citep{schad:2019}, especially for hierarchical models whose parameters have a complex relationship with the observed data. Standard prior elicitation strategies, such as \citet{ohagan:2004, moala:2010}, do not really help in such scenarios, since the expert still needs to express information in terms of probability distribution of the model's parameters. The idea of eliciting knowledge in terms of the observable data is not new -- in fact, it dates back to \citet{kadane:1980}. However, to our knowledge we proposed the first practical formulation that accounts for uncertainty in the expert's judgements of the prior predictive distribution, with easy, general, and complete implementation that allows eliciting both univariate and multivariate prior distributions more efficiently.

We demonstrated the general formalism in several practical contexts, ranging from simple conceptual illustrations and technical verifications to real elicitation examples. In particular, we showed that multivariate priors (of reasonable dimensionality) can be elicited in context of generalized linear models based on relatively small collection of probability judgements for different covariate sets. The approach can be coupled with existing modelling tools and used for eliciting prior information from real users, as demonstrated for the human growth model of \citet{preece:1978} implemented in \texttt{brms} \citep{burkner:2017}. Even though we only carried out a simplified and small-case experiment, the results already indicate that even users familiar with statistical modelling were more comfortable expressing knowledge of the observed data rather than model parameters, and that the resulting priors better matched their beliefs.

The obvious continuation of this work would consider tighter integration of the method into a principled Bayesian workflow, coupled with more extensive user studies. We also look forward to extend our method to cases of multiple experts opinions about the same observable variables. As a first attempt, we could consider the same predictive model and distinct $\alpha$'s for multiple experts. However, more work is needed in that regard.

\section*{Acknowledgements}

This work was supported by the Academy of Finland (Flagship programme: Finnish Center for Artificial Intelligence, FCAI; Grants 320181, 320182, 320183).

\bibliographystyle{rss}
\bibliography{../refs}

\newpage
\section*{SUPPLEMENTARY MATERIALS}

\subsection{Prior predictive probability}
In this section we highlight the steps to obtain the prior predictive probability, by rewriting it as a expected value w.r.t. the prior distribution as follows. Given that the probabilistic models, $\pi_{\Y|\btheta}(\y|\btheta)$ and the prior $\pi_{\btheta}$ are positive functions, we can rearrange the order of the integration \citep[See][Fubini-Tonelli theorem]{folland:2013}. Hence we have 

\begin{align} 
\mathbb{P}_{A| \blambda} &:= \displaystyle\int_A \pi_{\Y}(\y| \blambda) \ \mathrm{d} \hspace{-0.05cm} \y =  \displaystyle\int_A  \displaystyle\int_\Theta \pi_{\Y|\btheta}(\y|\btheta) \pi(\btheta| \blambda) \ \mathrm{d} \hspace{-0.05cm} \btheta \mathrm{d} \hspace{-0.05cm} \y \nonumber \\
& \stackrel{\textrm{Fub.}}{=} \displaystyle\int_\Theta \displaystyle\int_A \pi_{\Y|\btheta}(\y|\btheta) \pi(\btheta| \blambda) \ \mathrm{d} \hspace{-0.05cm} \y \mathrm{d} \hspace{-0.05cm} \btheta = \displaystyle\int_\Theta \mathbb{P}_{\Y|\btheta}({\Y \in A|\btheta}) \pi(\btheta| \blambda) \ \mathrm{d} \hspace{-0.05cm} \btheta \nonumber \\
&= \Ex_{\btheta}\big(\mathbb{P}_{\Y|\btheta}({\Y \in A|\btheta})\big).
\end{align}
%

\subsection{Approximate role of the precision measure}
Here we show the approximate behaviour of the precision parameter $\alpha$ for the general case when covariates are present. The simplification to other cases in straightforward. Recall the likelihood function of $\blambda$ given expert data reads,
\begin{align} \label{eq:sup_dir}
\mathcal{D}&(\pb_1, \ldots, \pb_J|\alpha, \blambda) = \dfrac{\Gamma(\alpha)^J} {\prod\limits_{j = 1}^J \prod\limits_{i_j = 1}^{n_j} \Gamma(\alpha \hspace{0.08cm} \mathbb{P}_{A_{j, i_j}| \blambda})} \prod_{j = 1}^J \prod_{i_j = 1}^{n_j} p_{j, i_j}^{\alpha \mathbb{P}_{A_{j, i_j}| \blambda} - 1}.
\end{align}
Consider the Stirling's approximation\footnote{This is a precise approximation.} to the $\Gamma(\cdot)$ function given by,
\begin{align}
\Gamma(x) \approx \sqrt{\dfrac{2\pi}{x}} \left(\dfrac{x}{e}\right)^x.
\end{align}

Rewriting the likelihood function in terms of the above approximation and removing terms that does not depend on $\alpha$ with 
a simplified notation we get, 
\begin{align}
\mathcal{D}(\pb|\alpha, \blambda) &\approx \dfrac{\left(\sqrt{\dfrac{2\pi}{\alpha}} \left(\dfrac{\alpha}{e}\right)^\alpha\right)^J} { \prod\limits_{j, i_j} \sqrt{\dfrac{2\pi}{\alpha \mathbb{P}_{A_{j, i_j}| \blambda}}} \left(\dfrac{\alpha \mathbb{P}_{A_{j, i_j}| \blambda}}{e}\right)^{\alpha \mathbb{P}_{A_{j, i_j}| \blambda}}} \exp\left(\sum_{i, j} \alpha (\mathbb{P}_{A_{j, i_j}| \blambda} - 1) \log p_{j, i_j} \right) \nonumber \\[0.6cm]
&\approx \dfrac{\alpha^{^{\sum_j n_j/2 - J/2}} \prod\limits_{i, j} \mathbb{P}_{A_{j, i_j}| \blambda}^{1/2}}{\exp\left(\alpha \sum\limits_{i, j}  \mathbb{P}_{A_{j, i_j}| \blambda} \log \ \dfrac{\mathbb{P}_{A_{j, i_j}| \blambda}}{p_{i, i_j}} \right)}
\end{align}
Take the logarithm of the above function and the derivative w.r.t. $\alpha$. Setting it to zero and solving for $\alpha$ we obtain,
\begin{align}
\hat{\alpha} \approx \dfrac{\sum_j n_j/2 - J/2}{\sum\limits_j KL(\mathbb{P}_{_j}||\pb_j)}
\end{align}
where the notation $\mathbb{P}_{_j} = [\mathbb{P}_{A_{j, 1}| \blambda} \cdots \mathbb{P}_{A_{j, n_j}| \blambda}]^\top$ and $KL(P||Q)$ denotes the Kullback-Leibler divergence in this order. 

\subsection{Hyperparameters' Fisher information matrix} 

The Fisher information matrix for the unknown hyperparameters can be obtained in closed-form by the fact that, in the original parametrisation of the Dirichlet distribution, the Fisher information is already known. In the original parametrisation and in its basic form, the probability density function reads
\begin{align}
\mathcal{D}(\pb|\alpha, \mathbb{P}) &= \dfrac{\Gamma(\alpha)}{\prod_{i = 1}^n \Gamma(\alpha \hspace{0.08cm} \mathbb{P}_i)} \prod_{i = 1}^n p_i^{\alpha \mathbb{P}_i -1}
\end{align}
where $\mathbb{P} = [\mathbb{P}_1 \cdots \mathbb{P}_n]^\top$.
Also knowing that the Dirichlet distribution belongs the exponential family, the Fisher information matrix reads,
\begin{align}
H_\mathbb{P} = \alpha^2 (\diag(\psi'(\alpha \mathbb{P})) - \psi'(\alpha) \mathds{1} \mathds{1}^\top),
\end{align}
whose inverse is given in closed-form as
\begin{align}
H_\mathbb{P}^{-1} = \tfrac{1}{\alpha^2}\bigg(\diag(\psi'(\alpha \mathbb{P}))^{-1} + \frac{\diag(\psi'(\alpha \mathbb{P}))^{-1} \mathds{1} \mathds{1}^\top \diag(\psi'(\alpha \mathbb{P}))^{-1}}{(1/\psi'(\alpha) - \mathds{1}^\top \diag(\psi'(\alpha \mathbb{P}))^{-1} \mathds{1})}\bigg)
\end{align}
where $\mathds{1}$ is $n \times 1$ vector with each component equals to $1$.

In the main paper, the vector of parameters $\mathbb{P}$ of the Dirichlet distribution is written as a function of $\blambda$. Using the change of variables for a new parametrisation \citep[see][page 64, Section 3.2.5, equation 3.27]{calder12, benmark11}, the Fisher information matrix with respect to $\blambda$ can be obtained directly (by passing any need of recalculating integrals) as, 
\begin{align} \label{eq:FIC1}
H_{\blambda} = \big[\tfrac{d}{d\lambda_1} \mathbb{P} \cdots \tfrac{d}{d\lambda_M} \mathbb{P}\big]^\top H_\mathbb{P} \big[\tfrac{d}{d\lambda_1} \mathbb{P} \cdots \tfrac{d}{d\lambda_M} \mathbb{P}\big]
\end{align}
where the vector $\tfrac{d}{d\lambda_m} \mathbb{P} = \big[\tfrac{d}{d\lambda_m}\mathbb{P}_1 \cdots \tfrac{d}{d\lambda_m}\mathbb{P}_n\big]^\top$ (the Jacobian matrix). Note that $H_\mathbb{P}$ is invertible and positive-definide, so as $H_{\blambda}$. Hence $H_{\blambda}$ is also invertible and its cholesky decomposition is stable to compute.

\paragraph{Presence of covariates (inputs):} When set of covariates are present, we have to consider that different partitions are provided. Since the likelihood function will still factorise for distinct covariates, note equation \eqref{eq:sup_dir}, the resulting Fisher information matrix will be the sum of Fisher information matrices \citep{casella:2001}. Hence, we can write, 
\begin{align} \label{eq:FIC2}
H_{\blambda} = \sum_j \big[\tfrac{d}{d\lambda_1} \mathbb{P}_j \cdots \tfrac{d}{d\lambda_M} \mathbb{P}_j \big]^\top H_{\mathbb{P}_j} \big[\tfrac{d}{d\lambda_1} \mathbb{P}_j \cdots \tfrac{d}{d\lambda_M} \mathbb{P}_j \big]
\end{align}

\subsection{Non-closed form prior predictive probabilities and hierachical structures}

For the case where $\mathbb{P}_j$ does not have closed-form expression we can estimate $\mathbb{P}_j$ and its derivatives w.r.t $\blambda$ using the {\it reparametrisation gradients} and automatic differentiation. The main idea is to find a pivotal function \citep[see][page 427, Section 9.2.2]{casella:2001} and obtain Monte-Carlo estimates of $\mathbb{P}_j$ and gradients $d/d \lambda_m \mathbb{P}_j$ with low computational cost according to \citet{figurnov:2018} and \citet{shakir:2019}. 

With a simplified notation, recall the prior distribution $\pi_{\btheta}$ and that the prior predictive probability can be rewritten as a expected value 
\begin{align}
\mathbb{P}_{A| \blambda} = \Ex_{\btheta}\big(\mathbb{P}({\Y \in A|\btheta})\big)
\end{align}
which depends on $\blambda$. Here the expression $\mathbb{P}({\Y \in A|\btheta})$ depends only on $\btheta$. Then, find a pivotal function $X = T(\btheta)$ such that the distribution of $X$ does not depend on $\blambda$. We then can rewrite the expectation,
\begin{align}
\mathbb{P}_{A| \blambda} = \Ex_X \big(\mathbb{P}({\Y \in A|T^{-1}_X(\blambda}))\big)
\end{align}
The gradients can be computed interchanging the order of integration and derivation,
\begin{align}
\dfrac{\mathrm{d}}{\mathrm{d} \lambda_m}\mathbb{P}_{A| \blambda} = \Ex_X \left( \dfrac{\mathrm{d}}{\mathrm{d} \lambda_m} \mathbb{P}({\Y \in A|T^{-1}_X(\blambda})) \right).
\end{align}
Where $T^{-1}_X(\cdot)$ is a inverse function of $T$ and depends on $X$ and $\blambda$.
The important notion here is that there is no need for resampling $X$ since the distribution $\pi_X(\cdot)$ is free of $\blambda$ by definition.

\paragraph{Hierachical structures:} Assume a hierarchical probabilistic model defined in form of layers as in the representation  $\Y \leftarrow \btheta_1 \leftarrow \cdots \leftarrow \btheta_L \leftarrow \blambda$, where the letter $L$ indicate the number of hierarchical layers. Formally one could write the hierarchical probabilistic model, 
\begin{align}
\Y|\btheta_1 &\sim \pi(\y|\btheta_1) \nonumber \\
\btheta_1|\btheta_2 &\sim \pi(\btheta_1|\btheta_2) \nonumber \\
&\vdots \nonumber \\
\btheta_L|\blambda &\sim \pi(\btheta_L|\blambda) 
\end{align}
whose prior predictive probability reads,
\begin{align}
\mathbb{P}_{A| \blambda} &= \displaystyle\int_\Theta \mathbb{P}({\Y \in A|\btheta_1}) \prod_{\ell = 1}^{L-1} \pi(\btheta_{\ell}|\btheta_{\ell + 1}) \pi(\btheta_L|\blambda) \mathrm{d} \hspace{-0.05cm} \btheta \nonumber\\
&=   \displaystyle\int_{\Theta_L}  \pi(\btheta_L|\blambda) \displaystyle\int_{\Theta_{L-1}}  \pi(\btheta_{L-1}|\btheta_L) \cdots  \displaystyle\int_{\Theta_1}  \pi(\btheta_1|\btheta_2) \mathbb{P}({\Y \in A|\btheta_1}) \mathrm{d} \hspace{-0.05cm} \btheta_1 \ldots \mathrm{d} \hspace{-0.05cm} \btheta_L
\end{align}
where $\Theta = \cup_{\ell = 1}^L \Theta_j$ and $\btheta_{\ell} \in \Theta_{\ell}$. Note that the above equation can be rewritten via the {\it tower property} by applying it sequentially due to the model hierarchy. 
\begin{align}
\mathbb{P}_{A| \blambda} = \Ex_{\btheta_L} \left( \Ex_{\btheta_{L-1}} \cdots \left(\Ex_{\btheta_1} \left( \mathbb{P}_{A|\btheta_1} \right) \right) \right) 
\end{align}
with shortened notation $\mathbb{P}({\Y \in A|\btheta_1}) = \mathbb{P}_{A|\btheta_1}$.
 
%
%
In this case, to apply the reparametrisation gradients technique, first find a pivotal function $X_\ell = T_j(\btheta_{\ell})$ for each layer $\ell$ whose inverse function is denoted as $\btheta_\ell$ $=$ $T^{-1}_{X_\ell}(\btheta_{\ell+1})$. 
Note the fact when we assume a pivotal quantity for every layer $\ell$, by definition the distribution of $\pi_{X_\ell}(x_\ell)$ $=$ $\pi_{\btheta_\ell|\btheta_{\ell+1}}(T^{-1}_{x_\ell})|\det J(T^{-1}_{x_\ell})|$ does not dependent on any $\btheta_{\ell + 1}$ or $\blambda$.
Hence, define the composite of inverse functions for each layer as $$\btheta_\ell = f_\ell(\blambda) = (T^{-1}_{X_\ell} \circ T^{-1}_{X_{\ell + 1}}  \circ \cdots \circ T^{-1}_{X_L})(\blambda)$$

This way, the above expected value as a function of $\blambda$ can be rewritten as, 
\begin{align} \label{eq:EXP}
\mathbb{P}_{A| \blambda} = \Ex_{X_L} \left( \Ex_{X_{L-1}} \cdots \left(\Ex_{X_1} \left( \mathbb{P}_{A|f_1(\blambda)} \right) \right) \right) 
\end{align}

To estimate $\mathbb{P}_{A| \blambda}$ via Monte Carlo first remember that $\blambda$ is fixed. Sample from $\pi_{X_\ell}$ for each $\ell$ and obtain the respectively the value of the function $f_\ell(\blambda)$ for each $\ell$. Calculate the sample mean of $\mathbb{P}_{A|f_1(\blambda)}$. Gradients of $\mathbb{P}_{A| \blambda}$  w.r.t. $\lambda$ can be obtained similarly, the extra step needed is in the calculation of the following expression, 
\begin{align} \label{eq:EXPG}
\dfrac{\mathrm{d}}{\mathrm{d}\lambda_m} \mathbb{P}_{A| \blambda} = \Ex_{X_L} \Bigg( \Ex_{X_{L-1}} \cdots  \left(\Ex_{X_1} \left(  \dfrac{\mathrm{d} f_1}{\mathrm{d} \lambda_m} \dfrac{\mathrm{d}} {\mathrm{d} f_1} \mathbb{P}_{A|f_1(\blambda)}  \right) \right) \hspace{-0.05cm} \Bigg) 
\end{align}
where the notation of the expectation $\Ex_{\boldsymbol{X}}(\cdot)$ is the same as in \eqref{eq:EXP}, but shortened. 
The first derivative on the right-hand side of the equation above then reads,
\begin{align} \label{eq:der1}
\dfrac{\mathrm{d} f_1}{\mathrm{d} \lambda_m} = \prod_{r = 1}^{L-1} \dfrac{\mathrm{d} T^{-1}_{X_r}}{\mathrm{d} T^{-1}_{X_{r+1}}} \dfrac{\mathrm{d} T^{-1}_{X_L}}{\mathrm{d} \lambda_m}.
\end{align}

In cases where the derivative of the inverse function $T_{X_\ell^{-1}}$  above cannot be obtained in closed-form we proceed similar as \cite{figurnov:2018} equation (6). Knowing that $T_\ell$ is one-to-one function, we can write
\begin{align}
X_{\ell} = T_{\ell}(T_{X_{\ell}}^{-1}(\btheta_{\ell+1}))
\end{align}
Take \textit{implicit} and \textit{explicit} derivatives (total derivative) with respect to $\btheta_{\ell +1}$ to get that 
\begin{align}
0 &=  \dfrac{\mathrm{d} T_{\ell}}{\mathrm{d}\btheta_{\ell + 1}}\bigg|_{\mathrm{explicit}} + \dfrac{\mathrm{d} T_{\ell}}{\mathrm{d}\btheta_{\ell + 1}}\bigg|_{\mathrm{implicit}} =  \dfrac{\mathrm{d} T_{\ell}}{\mathrm{d}\btheta_{\ell + 1}} +  \dfrac{\mathrm{d} T_{\ell}}{\mathrm{d}\btheta_{\ell}}  \dfrac{\mathrm{d} \btheta_{\ell}}{\mathrm{d}\btheta_{\ell + 1}} 
\end{align}
Identifying the notation $\btheta_{\ell} = T_{X_\ell}^{-1}$ for all $\ell$ and solving for $\dfrac{\mathrm{d} \btheta_{\ell}}{\mathrm{d}\btheta_{\ell + 1}}$ yields,
\begin{align} \label{eq:der2}
\dfrac{\mathrm{d} T^{-1}_{X_\ell}}{\mathrm{d} T^{-1}_{X_{\ell+1}}} = - \left(\dfrac{\mathrm{d} T_{\ell}}{\mathrm{d} T^{-1}_{X_{\ell}}} \right)^{-1} \dfrac{\mathrm{d} T_{\ell}}{\mathrm{d} T^{-1}_{X_{\ell+1}}} 
\end{align}
We can now plug \eqref{eq:der2} into \eqref{eq:der1} to estimate \eqref{eq:EXPG} and in turn to have the estimate for hyperparmeters' Fisher information matrix in \eqref{eq:FIC1} and \eqref{eq:FIC2}. Hence, we can proceed with {\it stochastic natural gradient descent} to estimate hyperparameters $\blambda$ for general types of probabilistic models.

\subsection{Predictive elicitation in practice: Example}

The probabilistic model for observed data (stature of male human being) is specified as follows, 
\begin{align}
Y_t|\btheta, b &\sim \mathcal{W}(h(t; \btheta), b) \nonumber \\ 
b &\sim \mathcal{G}(a_0, b_0) \nonumber \\
\theta_d &\stackrel{i.i.d}{\sim} \mathcal{LN}(a_d, b_d)
\end{align}
where $Y_t$ is univariate $S = 1$ and denotes the stature of the human being at time $t$. The parameters of the growth-model $h(t; \btheta)$ are denoted as $\btheta = [\theta_1 \ \theta_2 \ \theta_3 \ \theta_4 \ \theta_5] = [h_1, h_{t_*}, t_*, s_0, s_1]^\top$, where $h_1$ is the average height of an adult human, $h_{t_*}$ is the average high for the event "growth-spurt" \citep{preece:1978}, $t_*$ is when that event happens, $s_0$ and $s_1$ are constants from the model. The parameter $b$ controls the variance of the variable $Y_t$ around $h(t;\btheta)$. Large the values of $b$ less variance around the $h(t; \btheta)$ and vice-versa. $\mathcal{W}$, $\mathcal{G}$ and $\mathcal{LN}$ stands for respectively, Weibull, Gamma and log-Normal distributions.

We used the Weibull distribution in the mean-variance parametrisation which means that the probability distribution of $Y_t|\btheta, b$ is given by, 
\begin{align} \label{eq:weib}
\pi_{Y_t|\btheta, b}(y) = b \hspace{0.05cm} \tfrac{\Gamma(1 + 1/b)}{\exp\left(h(t;\btheta)\right)} \hspace{0.05cm} \left(y \hspace{0.05cm} \tfrac{\Gamma(1 + 1/b)}{\exp\left(h(t;\btheta)\right)} \right)^{b-1} \exp\left(- \left(y \hspace{0.05cm} \tfrac{\Gamma(1 + 1/b)}{\exp\left(h(t;\btheta)\right)} \right) ^b \right)
\end{align}
The other distribution used for the prior are used in their standard parametrisation scale-shape for Gamma and mean-variance for log-Normal distribution. 
The vector of hyperparameters is $\blambda = \{a_m, b_m, m = 0, \ldots, 5 \}$. The human-growth model obtained by \cite{preece:1978} and given in Section 2, Model 1 in their paper. In our notation this growth-model reads 
\begin{align} \label{eq:sol}
h(t; \btheta) = h_1 - \dfrac{2(h_1 - h_{t_*})}{\exp[s_0 \ (t-t_*)] + \exp[s_1 \ (t-t_*)]}.
\end{align}

The only general background information provided to the participants was the following brief description characterizing the overall growth process and providing general numerical values as reminders:\\

\textit{"During the early stages of life the stature of female and male are about the same, but their stature start to clearly to differ during growth and in the later stages of life. In the early stage man and female are born roughly with the same stature, around $45$cm - $55$cm. By the time they are born reaching around 2.5 years old, both male and female present the highest growth rate (centimetres pey year). It is the time they grow the fastest. During this period, man has higher growth rate compared to female. For both male and female there is a spurt growth in the pre-adulthood. For man, this phase shows fast growth rate varying in between 13-17 years old and female varying from 11-15. Also, male tend to keep growing with roughly constant rate until the age of 17-18, while female with until the age of 15-16. After this period of life they tend to stablish their statures mostly around $162$ - $190$cm and $155$ - $178$cm respectively."} \\

Given the background information we asked each user to provide the distribution for statures of males at given ages $t = \{t_1, t_2, t_3, t_4\} = \{0, 2.5, 10, 17.5 \}$ in form of probabilistic assessments. For eliciting the probabilities we asked them to provide the thresholds $y_i$ determining the statures that partition the sample space with the following probabilities
%
\begin{align}
\mathbb{P}(Y_t \leq y_1) = 0.10 \nonumber \\
\mathbb{P}(Y_t \leq y_2) = 0.25 \nonumber \\
\mathbb{P}(Y_t \leq y_3) = 0.50  \nonumber \\
\mathbb{P}(Y_t \leq y_4) = 0.75  \nonumber \\
\mathbb{P}(Y_t \leq y_5) = 0.90   
\end{align}
where naturally $y_1 < y_2 < \ldots < y_5$. The data used as each $t_j$ was hence given by
\begin{align}
\mathbb{P}(Y_{t_j} \in (0, y_1)) &= p_{j, i_j} = 0.10   \nonumber \\
\mathbb{P}(Y_{t_j} \in (y_1, y_2)) &= p_{j, i_j} = 0.15 \nonumber \\
\mathbb{P}(Y_{t_j} \in (y_2, y_3)) &= p_{j, i_j} = 0.25 \nonumber \\
\mathbb{P}(Y_{t_j} \in (y_3, y_4)) &= p_{j, i_j} = 0.25 \nonumber \\
\mathbb{P}(Y_{t_j} \in (y_4, y_5)) &= p_{j, i_j} = 0.15 \nonumber \\
\mathbb{P}(Y_{t_j} \in (y_5, \infty)) &= p_{j, i_j} = 0.1  
\end{align}
%

\subsection*{Results for the prior predictive elicitation} 
The main manuscript provided the results for one example user. The results for other four users are provided here in Tables~1 to 4.
The general trend of prior predictive elicitation matching better the data-dependent values of \citet{preece:1978} remains, and for some users the direct parameter elicitation approach resulted in very poor prior (e.g. $h_{t_*}$ for User 3).
\\
\begin{table}
\caption{User 2} 
\begin{tabular}{crrrrr}
       &           & \multicolumn{2}{c}{\textbf{Predictive}} & \multicolumn{2}{c}{\textbf{Parametric}} \\ 
Parameter   & Reference & $\Ex[\cdot]$ & $\Vx(\cdot)$ & $\Ex[\cdot]$ & $\Vx(\cdot)$ \\ \hline
$h_1$ & 174.6  & 191.74 & 4.32             & 172.7 & 101.6 \\
$h_{t_*}$     & 162.9   & 153.73           & 1.6             & 129.1 & 31.0  \\
$s_0$ & 0.1   & 0.04    & $<$ 0.01 & 0.51  & $<$ 0.04   \\
$s_1$ & 1.2   & 2       & 4.3              & 0.5   & $<$ 0.04  \\
$t_*$ & 14.6  & 15.9    & 0.7              & 12.9  & 0.5  \\
$b$  &        & 61.4    & 111.4            & 3.1   & 2.6  \\
\hline
$\alpha$    & $-$          & 14.0   & $-$               &  1.3 & $-$                
\end{tabular}%
\end{table}
\begin{table}
\caption{User 3}
\begin{tabular}{crrrrr}
       &           & \multicolumn{2}{c}{\textbf{Predictive}} & \multicolumn{2}{c}{\textbf{Parametric}} \\ 
Parameter   & Reference & $\Ex[\cdot]$ & $\Vx(\cdot)$ & $\Ex[\cdot]$ & $\Vx(\cdot)$ \\ \hline
$h_1$     & 174.6 & 177.14 & 3.68            & 174.6 & 146.3 \\
$h_{t_*}$ & 163.0 & 148.8  & 1.86            & 78.5  & 37.2 \\
$s_0$     & 0.1   & 0.07   & $<$ 0.001        & 0.2   & 0.004 \\
$s_1$     & 1.2   & 4.54   & 37.83           & 0.9   & 0.004 \\
$t_*$     & 14.6  & 11.31  & 0.21            & 6.9   & 2.9  \\
$b$       &   $-$    & 18.4  & 12.5           & 25.8  & 74.1 \\
\hline
$\alpha$    & $-$          &  9.5    & $-$               & 1.5   		& $-$                
\end{tabular}%
\end{table}
\newpage
\begin{table}
\caption{User 4}
\begin{tabular}{crrrrr}
       &           & \multicolumn{2}{c}{\textbf{Predictive}} & \multicolumn{2}{c}{\textbf{Parametric}} \\ 
Parameter   & Reference & $\Ex[\cdot]$ & $\Vx(\cdot)$ & $\Ex[\cdot]$ & $\Vx(\cdot)$ \\ \hline
$h_1$     & 174.6  & 174.5    & $<$ 0.01    & 50.5  & 64.5 \\
$h_{t_*}$ & 162.9  & 162.8    & 0.02        & 129.1 & 31.0 \\
$s_0$     & 0.1    & 0.1     & $<$ 0.01     & 5.1   & 2.7  \\
$s_1$     & 1.2    & 1.6 & 1.7   & 5.1   & 2.7  \\
$t_*$     & 14.60  & 14.7    & 0.9          & 12.9  & 0.6  \\
$b$       &   $-$     & 14.5    & 14.3         & 1      & $<$ 0.02  \\
\hline
$\alpha$    & $-$          & 17.1	   & $-$  &   1.2	 & $-$                
\end{tabular}%
\end{table}
\begin{table}
\caption{User 5}
\begin{tabular}{crrrrr}
       &           & \multicolumn{2}{c}{\textbf{Predictive}} & \multicolumn{2}{c}{\textbf{Parametric}} \\ 
Parameter   & Reference & $\Ex[\cdot]$ & $\Vx(\cdot)$ & $\Ex[\cdot]$ & $\Vx(\cdot)$ \\ \hline
$h_1$     & 174.6   & 174.4 & 0.91            & 159.66 & 155.96 \\
$h_{t_*}$ & 162.9   & 162.6 & 0.85            & 121.75 & 57.27  \\
$s_0$     & 0.1     & 0.1   & $<$ 0.01 & 3.3      & 3.3      \\
$s_1$     & 1.2     & 3.4   & $<$ 0.01 & 3.3      & 3.3      \\
$t_*$     & 14.6    & 14.6  & 0.02            & 11.7   & 5.36   \\
$b$       & $-$     & 17.8   & 17.8            & 9.5      & 8.3     \\
\hline
$\alpha$    & $-$          & 7.7		   & $-$               & 1.5    		& $-$                
\end{tabular}%
\end{table}

\end{document}